# THE OPTIMAL $N$-BODY METHOD FOR STABILITY STUDIES OF GALAXIES

DAVID J. D. EARN

Institute of Astronomy, Madingley Road, Cambridge CB3 0HA, UK,
and Racah Institute of Physics, The Hebrew University, Jerusalem 91904, Israel*: earn@astro.huji.ac.il

AND

J. A. SELLWOOD

Department of Physics and Astronomy, Rutgers University, Piscataway, NJ 08855: sellwood@physics.rutgers.edu

## ABSTRACT

The stability of a galaxy model is most easily assessed through $N$-body simulation. Particle-mesh codes have been widely used for this purpose, since they enable the largest numbers of particles to be employed. We show that the functional expansion technique, originally proposed by Clutton-Brock for other simulation problems, is in fact superior for stability work. For simulations of linear evolution it is not much slower than grid methods using the same number of particles, and reproduces analytical results with much greater accuracy. This success rests on its ability to represent global modes with a modest number of basis functions; grid methods may be more effective for other applications, however. Our conclusions are based on implementations of functional expansion and grid algorithms for disk galaxies.

*Subject headings*: galaxies: kinematics and dynamics — galaxies: structure — instabilities — methods: numerical — celestial mechanics: stellar dynamics

## 1. INTRODUCTION

$N$-body simulations have been used for a number of years to mimic the dynamical behavior of collisionless stellar systems. When the number of particles employed is many orders of magnitude fewer than the number of stars in a galaxy, the level of shot noise in the density distribution is greatly enhanced. Potential fluctuations arising solely from particle noise cause the system to relax and determine the time-scale over which the simulation approximates the collisionless limit.

Thus all simulation techniques (see Sellwood 1987 for a review) seek a potential that arises from a density function that is smoother than the particle distribution. An ideal smoothing procedure would retain meaningful features of the particle distribution while suppressing variations in density that arise solely from noise. The smoothing problem also arises in density estimation from noisy experimental data, where standard techniques have been grouped into three classes (e.g., Silverman 1986; Merritt & Tremblay 1994):

The first, and conceptually the simplest, method is to use a local smoothing kernel, which replaces each particle by an extended, usually spherical, cloud of matter. The effect in an $N$-body simulation is to introduce a short-range cut-off in the interparticle forces, known as softening. It is inserted explicitly into direct codes (e.g., Aarseth 1985) or tree codes (e.g., Barnes & Hut 1986; Hernquist 1987), known as "Particle-Particle" (PP) methods, and implicitly in grid codes, known as "Particle-Mesh" (PM) methods. The smoothed density profile is biased, however, because it can never be more peaked than the width of the kernel. Reducing the kernel width reduces the bias, but results in a noisier function – i.e., one having a larger variance. Variable softening (White 1982) or grid cell size (van Albada 1982) can lessen this dilemma in 3-D systems with large density variations, but use of a fully adaptive kernel (Merritt & Tremblay) in an $N$-body simulation would substantially increase the calculation time (and e.g., would violate Newton's third law).

---

* Current address.

The second technique expands the particle distribution in a set of functions; the small scale density variations are filtered out by truncating the expansion after a few terms. This is essentially a parametric method since the density estimate is biased by the adopted basis, which reflects a prejudice about the form of the true function. The quality of the approximation to the true density depends, therefore, upon how well the low order members of the chosen set match the profile. Attempting to compensate for a bad choice of basis by adding more functions again trades reduced bias for increased variance. Likewise, cranking up the number of particles reduces the variance, but clearly does not lessen the bias.

Strictly speaking, an expansion in smooth functions can be regarded as replacing each point mass by a complicated (non-local) smoothing kernel, although this viewpoint seems quite contrived because the shape of the implied mass cloud depends on the position of the particle relative to the expansion center. We feel it is more useful to distinguish the expansion approach, which cannot be cast as a convolution, as filtering out the high spatial frequencies, and softening as a convolution with a local smoothing kernel.

A third (non-parametric) approach, favored by Merritt & Tremblay (1994) for data analysis, is the maximum penalized likelihood method. They describe how a one-dimensional $N$-body code could be devised using this method; it would compute a so-called "smoothing spline" estimate of the interior mass using a penalty functional that guarantees smoothness of the solution. Choosing this penalty functional to make the resulting density a piecewise polynomial (Wahba 1990) would be particularly efficient, but other choices might be more appropriate for computation of the radial force. Generalization of this idea to higher dimensions has yet to be explored, however.

It is generally agreed (e.g., Spitzer & Hart 1971; Rybicki 1972) that local smoothing inhibits two-body relaxation – the most extreme consequence of particle noise on short scales. On the other hand, large-scale density fluctuations caused by shot noise cannot be distinguished from physically meaningful variations on the same scale and none of the above methods can prevent it from influencing the behavior of the system. For some applications, the effects of noise on large scales can be controlled by a quiet start (Sellwood 1983) in which the particles are placed symmetrically in azimuth, and wavenumbers affected by the periodicity (either directly or through aliases) are removed by filtering. [Similar techniques are standard in cosmological $N$-body simulations (e.g., Efstathiou *et al.* 1985).] Failing this, the problem can be combatted by increasing the number of particles, a strategy that yields slow returns and places a high premium on numerical efficiency.

Of the kernel methods, grid codes (Miller & Prendergast 1968; Hockney & Hohl 1969; Hockney & Brownrigg 1974; Miller 1976; James & Sellwood 1978; Miller & Smith 1978; Hohl & Zang 1980; Sellwood 1981; van Albada 1982; Pfenniger & Friedli 1993) have proved the most successful for isolated galaxies because they enable the largest numbers of particles to be employed. Their principal disadvantage is the lack of versatility (or bias) imposed by a fixed grid shape and spatial resolution, as well as a finite calculation volume. However, these are less serious handicaps when studying the comparatively mild evolution of near equilibrium systems, especially when the grid geometry is tailored to the problem. The much slower PP methods can never employ the numbers of particles possible with PM methods, but being more generic, they are useful for situations where no grid code has yet been developed; examples are systems with multiple centers, such as galaxy groups, or cases where the mass distribution undergoes wholesale rearrangement, e.g., galaxy mergers.

One example of the serious consequences of bias in kernel methods arises in simulations of a stellar disk; the growth rates of instabilities are reduced when the softening length is comparable to, or greater than, the disk thickness. Quite modest values can entirely suppress even one of the largest scale instabilities of disks – the bar instability (Sellwood 1981, 1983). This particular problem is avoided in an expansion method: the particle distribution yields a direct estimate of the amplitude of the large-scale density variation that is not biased by having been first convolved with a local smoothing kernel. To clarify the essential difference here, we note that softened gravity of the standard form $[\Phi = -G(r^2 + \epsilon^2)^{-1/2}]$ is well known to yield a gravitational potential in a thin disk that is equal to the Newtonian value in a plane offset vertically by a distance $\epsilon$. Since the potential of a wafer-thin mass distribution having a Bessel function form decays as $\exp(-k|z|)$ away from the plane, we could readily mimic softening using these functions by simply reducing the potential of each by the factor $\exp(-k\epsilon)$. This demonstrates that the standard softening kernel weakens the disturbance force even from smooth, large-scale density perturbations, whereas Bessel functions (or any other basis set) yield the Newtonian field. The potential from a truncated expansion is still biased, of course: disturbances whose wave numbers are omitted from the truncated basis are obviously excluded, and the spatial forms of the density variations may be inadequately represented by an inappropriate, or too drastically truncated, basis.

The expansion technique was originally pioneered by Clutton-Brock (1972; 1973), and has recently undergone something of a revival (Allen, Palmer & Papaloizou 1990; Hernquist & Ostriker 1992). It is nearly as efficient as grid methods, although it may be still less versatile. The potential is the exact solution of Poisson's equation for the spatially filtered density distribution given by the first few terms of the series of selected basis functions. This can be viewed as a Monte Carlo method, the particles being a representative sample of those in the true system that is used to determine the coefficients of the density expansion. Conceptually at least, particles do not interact in a pairwise manner but each with the large-scale force field of the galaxy, and the technique has therefore been expected to be "less collisional" than kernel methods (e.g., Allen *et al.* 1990). To reflect this conceptual difference from PP and PM methods, we refer to basis expansion as a "Smooth-Field-Particle" (SFP) technique. (Hernquist & Ostriker used the acronym SCF for "Self-Consistent Field" but we prefer SFP because it emphasizes that the technique is based upon particles.)

However, Hernquist & Ostriker reported that the energy fluctuations of individual particles in a stable spherical model were scarcely any less when the evolution was computed by their newly developed SFP code than by tree or other algorithms. This finding seemingly reinforced the conclusion already reached by Hernquist & Barnes (1990) that the relaxation rate in any valid $N$-body technique depends only on $N$ and is essentially independent of the smoothing method.

Their results re-emphasize the detrimental effects of shot noise. Sellwood (1983) found that the dominant instability of a stellar disk stood out clearly when he used a quiet start but he also reported that interference by many other large-scale modes seeded with a high initial amplitude introduced very large uncertainties in noisy start models. Weinberg (1993) pointed out that the energy fluctuations observed by Hernquist and his collaborators probably resulted from large-scale modes, especially stable oscillations, excited by the random selection of a finite number of particles to represent a smooth density distribution.

The apparent conclusion that no valid $N$-body method is superior to any other is therefore no more than a reflection of the uniform influence of noise on large scales when no steps are taken to suppress it. For studies of the stability of equilibria, we show in this paper that the SFP technique *is* superior to a grid code, and presumably to other techniques, once particle noise has been controlled by a quiet start. The biases in the SFP method are ideally suited to select for the linear behavior, which we show can be reproduced almost perfectly while avoiding the systematic errors introduced by a softening kernel.

## 2. TECHNIQUE

The principles of the SFP method have been described both by Clutton-Brock (1972) and by Hernquist & Ostriker (1992); however, we feel it useful to present a brief, and somewhat distinct, outline here.

### 2.1. *The Smooth-Field-Particle Approach*

In order that we can evaluate the force field directly from the smoothed density given by a truncated series expansion, we require a set of function *pairs* $\{\rho_j, \psi_j\}$ such that

$$\psi_j(\boldsymbol{x}) = -G \int \frac{\rho_j(\boldsymbol{x}')}{|\boldsymbol{x}' - \boldsymbol{x}|} d^3\boldsymbol{x}', \qquad j = 0, \ldots, \infty. \tag{1}$$

These basis functions need not be realistic density-potential pairs themselves – any pairs of (complex) functions that formally solve (1) may be used.

The inner product of two density or potential functions is defined by minus the interaction potential energy of the two disturbances with densities $\rho$ and $\rho'$,

$$\begin{aligned} \langle \rho, \psi' \rangle &\equiv -\frac{r_0}{GM^2} \int \rho^* \psi' \, dV \\ &= -\frac{r_0}{4\pi G^2 M^2} \int (\nabla^2 \psi^*) \psi' \, dV \ . \end{aligned} \tag{2}$$

Here, the asterisk denotes complex conjugation and the factor $r_0/GM^2$ (where $r_0$ and $M$ are length and mass scales) is included so that the inner product is dimensionless. To put the theory on a rigorous foundation (e.g., Kalnajs 1971) it is necessary to restrict attention to density and potential functions with a finite norm ($\|\rho\| = \|\psi\| \equiv \langle \rho, \psi \rangle^{1/2}$) i.e., mass distributions with finite potential energy.

Since we need to be able to approximate the potential from any distortion of our initial model, we require any basis set that we use to be complete in some sense. Normally this means that for any mass distribution with finite potential energy the series

$$\rho = \sum_{j=0}^{\infty} c_j \rho_j \ , \qquad \psi = \sum_{j=0}^{\infty} c_j \psi_j \ , \tag{3}$$

converge in the mean (e.g., Arfken 1985, p. 524). A weaker, and probably adequate form of completeness is discussed by Saha (1993).

A basis set is said to be *biorthogonal* if $\langle \rho_j, \psi_k \rangle \neq 0$ if and only if $j = k$, and *biorthonormal* if

$$\langle \rho_j, \psi_k \rangle = \delta_{jk}, \qquad \text{for all } j \text{ and } k. \tag{4}$$

With a biorthonormal basis, the dimensionless expansion coefficients are simply $c_j = \langle \rho, \psi_j \rangle = \langle \rho_j, \psi \rangle$. Any basis can be made biorthonormal with the Gram-Schmidt algorithm (e.g., Arfken 1985).

In an $N$-body experiment, a galaxy is modelled by $N$ point particles with masses $\{M_i\}$ and positions $\{\boldsymbol{x}_i(t)\}$, so the full density of the galaxy at time $t$ is

$$\rho(\boldsymbol{x}, t) = \sum_{i=1}^{N} M_i \, \delta(\boldsymbol{x} - \boldsymbol{x}_i(t)) \ . \tag{5}$$

Using basis density functions $\{\rho_j\}$ we may also write

$$\rho(\boldsymbol{x},t) = \sum_{j=0}^{\infty} c_j(t)\rho_j(\boldsymbol{x}) . \tag{6}$$

The essence of the SFP method is to make the approximation

$$\rho(\boldsymbol{x},t) \simeq \sum_{j=0}^{j_{\max}} c_j(t)\rho_j(\boldsymbol{x}) , \tag{7}$$

with small $j_{\max}$. If the basis is biorthonormal, it follows that

$$\begin{aligned} c_j(t) &= \langle \rho(\boldsymbol{x},t), \psi_j(\boldsymbol{x}) \rangle \\ &= -\frac{r_0}{GM^2} \sum_{i=1}^{N} M_i \psi_j(\boldsymbol{x}_i(t)) . \end{aligned} \tag{8}$$

For a non-biorthogonal basis the potential coefficients are obtained from the quantities in (8) by multiplying by a $j_{\max} \times j_{\max}$ matrix (e.g., Saha 1993). Since the matrix elements can be computed once in advance, the extra expense due to non-biorthogonality is negligible provided $j_{\max}^2 \ll N$.

In principle, (5) violates the condition that our mass distributions must generate finite potential energies. In practice, however, delta functions need not be used in (5) to obtain (8). If finite size particles are used then (8) will be approximate, but as the particles are reduced in size the magnitude of the error can be reduced below the round-off error of the computer used.

The exact Newtonian force field generated by the truncated density expansion is easily computed from the coefficients. The force on particle $i$ at time $t$ is

$$\boldsymbol{F}_i(t) = -M_i \sum_{j=0}^{j_{\max}} c_j(t) \nabla \psi_j(\boldsymbol{x}_i(t)) . \tag{9}$$

The gradient of the basis potentials can be calculated analytically so no further approximation is made here.

It is worth highlighting a numerical consideration at this point. All but the lowest order monopole functions of any reasonable basis set are oscillatory, and the summation (8) will contain many cancelling contributions resulting in significant loss of precision, which worsens as the number of particles is increased. It is therefore essential to compute the functions and the sum (8) using double precision arithmetic when employing more than a few hundred particles [though the forces (9) can be calculated in single precision].

### 2.2. *Choice of basis functions*

Our first implementation of this method has been an application to flat disk galaxies. For this problem it is most natural to use potential and surface density functions that separate in polar coordinates,

$$\psi_{mn}(r,\theta) = e^{im\theta}\phi_{mn}(r) , \qquad \Sigma_{mn}(r,\theta) = e^{im\theta}\sigma_{mn}(r) . \tag{10}$$

For most applications it is preferable to choose the radial functions to have infinite extent because the system has no natural outer boundary, even if it is initially confined to a finite region. Standard bases having this property and that could be used for flat disks include Bessel functions, logarithmic spirals (Kalnajs 1971), those given by Clutton-Brock (1972) and Qian (1993) and the complete but non-biorthogonal sets suggested by Qian (1992).

However, for linear stability studies of even an infinite axisymmetric equilibrium system, a basis set of finite radial extent can be used since the instabilities are expected to be confined to the inner parts. Moreover, for this restricted application, we can exploit the fact that different azimuthal components of the density are decoupled at small amplitude – the usual result from linear theory. Simulations can therefore be performed with a single active azimuthal harmonic ($e^{im\theta}$, $m \neq 0$) with the unperturbed axisymmetric force treated as a fixed background field. (A little more care is needed when using a finite basis for axisymmetric oscillations in an infinite disk since only part of the mass can be active. The central attraction must be divided into the active part that comes from the particles and a rigid supplementary force from the inactive mass, which is needed to maintain the radial balance.)

Thus for the problem at hand, we merely require that the first few members of the basis set chosen are able to represent the low-order growing modes without undue bias. As the radial parts of the non-axisymmetric functions in a typical basis set are similar in form to the axisymmetric functions, it may be advantageous to choose a basis that can represent the unperturbed model with a small number of functions. There is no real advantage, however, to insisting that the lowest order $m = 0$ basis function has the density profile of the initial configuration, since the non-axisymmetric modes will almost certainly not be single functions of the same basis. The Maclaurin disk provides an extreme example: the surface density declines as $[1 - (r/r_0)^2]^{1/2}$ while the normal modes (Hunter 1963; Kalnajs 1972) are singular at the edge and bear no resemblance to the unperturbed surface density.

Here we adopt the Abel-Jacobi basis sets (given in the Appendix) which were obtained by Kalnajs (1976a) for flat disks. The potential functions, $\phi_{mn}(r)$, in each of these bases are polynomials in $r$ and the surface densities, $\sigma_{mn}(r)$, are polynomials times a factor $[1 - (r/r_0)^2]^{k-1/2}$ where $r_0$ is the "function edge", the radial limit of the functions. Each choice of the non-negative integer $k$ specifies a different basis set, and each member of a set of given $k$ is specified by the number of azimuthal nodes $m$ and the number of radial nodes $n$. The $\sigma_{00}$ term corresponds to a surface density that ranges from one that is singular at the outer edge for $k = 0$, to a point mass as $k \to \infty$. Intermediate values of $k$ are useful for realistic finite disk galaxy models.

### 2.3. *Efficiency*

Since the quality of an $N$-body code is fundamentally limited by the number of particles, the viability of a technique is determined to a large extent by its speed. The computer time required for an SFP simulation depends linearly on the product of the particle number $N$ and the number of basis functions. This weak $N$-dependence is greatly superior to all PP methods (direct or tree) and is rivalled only by PM methods.

The computational expense of calculating the coefficients (8) clearly depends strongly on the complexity of the basis functions. But all basis sets become equally efficient if a table of function values is stored rather than evaluating the exact functions every time they are required. We create tables of $10^4$ values for the radial part of each basis function employed and adopt a fourth-order (five point) Lagrange interpolation formula which yields values good to about 15 figures from just nine arithmetic operations per evaluation. (Recall that the functions need to be evaluated in double precision.) The time required to make the tables is negligible but reduces the overall running time of the code by an order of magnitude. We find that for a given number of particles, the SFP code (using 13 functions) runs about three times more slowly than the 2-D polar PM code (with a mesh size of $85 \times 128$). When more than a single azimuthal wavenumber is active, it is clearly advantageous to group contributions with the same $m$ to save repeated evaluations of identical trigonometric factors.

An important feature of grid methods is that they are very well suited to vector and parallel machines. This is equally true for the SFP algorithm, as has been emphasized by Hernquist & Ostriker (1992).

### 2.4. *Making the Best Use of Each Particle*

There are two separate parts to our procedure for reducing noise in the initial particle distribution. *N.B.*, these techniques are independent of the method of force determination used in the simulation.

Conceptually the simplest, and the easiest to implement, is to reduce non-axisymmetric density variations by placing several particles having identical radial and azimuthal velocities at equal angular intervals at the same radius. This is very effective even for PP or Cartesian PM methods for which filtering of azimuthal wavenumbers cannot be implemented without considerable extra effort. We can do even better for polar PM and SFP methods because the force calculation is readily restricted to azimuthal wavenumbers up to $m_{max}$, say, in which case we require $2(m_{max}+1)$ equally spaced particles to mimic an exactly smooth ring. [The factor 2 is required to prevent interference from aliases (Sellwood 1987).] If odd wavenumbers are excluded, then we need only a half-ring, *etc.* In the calculations presented here, non-axisymmetric forces are restricted to $m=2$; we therefore need place just three particles on each half-ring in order to reduce the initial amplitude of $m=2$ disturbances to very low values, and allow us to observe linear growth over many $e$-folds.

A further substantial improvement results from selecting particles from the DF so as to reduce sampling noise. The density of a sample of particles selected at random from the DF will inevitably possess random fluctuations about the desired DF. The dynamical behavior of the $N$-body model will mimic that of a smooth system having this somewhat different DF, a difference that is easily detectable: the eigenfrequency of the dominant mode varies randomly for different random selections of particles while all other numerical parameters are held fixed. These random variations could, of course, be reduced by increasing the number of particles, but they can be virtually eliminated by adhering to the almost fully deterministic procedure for selecting particles described by Sellwood & Athanassoula (1986). Since the DF is a function only of the isolating integrals, we do not need to select particles smoothly in full phase space but merely in the sub-space of the integrals. The procedure is analgous to the optimal choice of abscissae in multi-dimensional Monte Carlo integration (e.g., Press *et al.* 1992, §§7.6–7.8). We integrate the DF to obtain a mass function, and place abscissae at equal mass increments; this procedure yields deterministically distributed abscissae, spaced sparsely where the DF has low values and densely where it is large. Each selected abscissa (point in integral space) defines an orbit on which a (constant) number of particles must be positioned. Results for non-axisymmetric modes are quite insensitive to choices of the radial phase on this orbit, whereas a smooth distribution of azimuthal phases, as described above, is essential. The efficacy of this procedure is evident from the results presented in §3.2.

### 3. EIGENFREQUENCY TEST

### 3.1. *Background*

Kalnajs's study of the isochrone disks is one of the few to have considered the normal modes of realistic stellar disks with velocity dispersion. [Hunter's (1992) study of the Kuz'min-Toomre models is another example.] The Maclaurin disk, solved completely by Kalnajs (1972), is uniformly rotating and quite unlike a galaxy [though Sellwood (1983) found it afforded a useful test]; the models studied by Sawamura (1988) are not much more realistic in this respect, while the $V_c =$ constant disk studied by Zang and Toomre (Zang 1976; Toomre 1981) presents significant numerical difficulties – we will report on simulations of this model in a future paper. Toomre's other work (e.g., Toomre 1981) has focused on models with stars on initially circular orbits, using softened gravity to suppress axisymmetric Jeans instabilities.

The surface density of the flat isochrone disk is

$$\Sigma(r) = \frac{Ma}{2\pi r^3} \left[ \ln\left(\frac{r + \sqrt{a^2 + r^2}}{a}\right) - \frac{r}{\sqrt{a^2 + r^2}} \right] , \qquad (11)$$

which gives rise to the potential in the disk plane

$$\psi(r) = -\frac{GM}{a + \sqrt{a^2 + r^2}} . \qquad (12)$$

Evans and de Zeeuw (1992) give the full three-dimensional potential. Kalnajs (1976b) gave a family of DFs for the flat isochrone disk, which are characterised by a single parameter $m_{\rm K}$ that determines the degree of radial pressure; the models are denoted "isochrone/$m_{\rm K}$" (higher values of $m_{\rm K}$ yield a cooler disk).

We concentrate here on the isochrone/12 model, but we have also run experiments with $m_{\rm K} = 9$ and $m_{\rm K} = 6$; all these models are axisymmetrically stable – Toomre's $Q$ is roughly constant at about unity for the isochrone/12 model. Kalnajs (1978) reported the eigenfrequencies of the dominant bisymmetric eigenmodes of the isochrone/$m_{\rm K}$ disks for several $m_{\rm K}$. His brief paper gives few details; in particular, he did not describe the rule he adopted to introduce retrograde stars. The DF he used was

$$F'(J_r, J_\theta) = \begin{cases} \frac{1}{2} F(J_r + |J_\theta|, 0), & \text{if } J_\theta < 0, \\ F(J_r, J_\theta) - \frac{1}{2} F(J_r + |J_\theta|, 0) & \text{if } J_\theta > 0, \end{cases} \qquad (13)$$

where $F$ denotes the isochrone/$m_{\rm K}$ DF for direct stars only, and $J_r$ and $J_\theta$ are the radial action and angular momentum respectively (Kalnajs, private communication).

Following Kalnajs (1978), we truncate the active disk by excluding particles having sufficient energy for any part of their orbits to cross an outer limiting radius, which we choose to be at $r_0 = 5a$. For a disk with some random motion, the surface density of the surviving particles tapers smoothly to zero at $r_0$.

Zang & Hohl (1978) have already reported simulations of isochrone disks on a Cartesian grid, but without using a quiet start and making no corrections for softening and other grid effects. Their estimates of the growth rates seemed to agree with Kalnajs's predictions at the $\sim 20\%$ level. Here we report that simulations with a quiet start (Sellwood 1983) performed with our SFP code can reproduce the predicted eigenfrequency to at least the precision of the prediction. With the polar grid code, on the other hand, we have to perform simulations having a range of softening lengths and extrapolate the measured eigenfrequencies to zero softening obtaining only approximate agreement. With both codes, we find a second $m = 2$ mode, not reported by Kalnajs, which grows almost as vigorously as the dominant mode.

We have used the mode fitting procedure devised by Sellwood and Athanassoula (1986) to derive the eigenfrequencies from the simulation. The procedure is illustrated in Figure 1. Briefly, the input data from the SFP runs are the sequence of Abel-Jacobi expansion coefficients $c_{mn}$ for a given $m$, while those from the grid code are from an expansion in logarithmic spirals. To these data, we fit one or more (two in the case of Figure 1) growing and rotating linear combinations of the expansion functions and minimize the squared residuals as the growth rates and pattern speeds are varied. We estimate the error in the eigenfrequencies by repeating the procedure many (typically 20) times using various subsets of the data and different weights for different time ranges (see Sellwood and Athanassoula for details).

### 3.2. SFP *code results*

Figure 2 shows how the estimated eigenfrequencies of the two most unstable modes vary with the number of ($m = 2$) basis functions used, for both the $k = 2$ and $k = 7$ Abel-Jacobi sets. Kalnajs's prediction for the dominant mode was obtained using his matrix method (Kalnajs 1977) employing the $k = 7$ basis and $n_{\max} = 15$. Our results with this same basis confirm that the frequency converges to the same value $(0.59 + 0.21i)$ when $n_{\max} \gtrsim 12$. Results with the $k = 2$ basis also appear to converge to the same frequency, but at a larger $n_{\max}$.

Figure 3 shows that our estimated eigenfrequency for the dominant mode varies very little as the number of particles is increased by a factor of 32. It differs from the predicted value by no more than 5% with as few as 15K particles, but the rate of convergence to the continuum value from this excellent start is very slow because our set-up procedure (§2.4) is so successful at mimicking a smooth distribution with the available particles. The weak $N$-dependence reflects the slow improvement with sample size in the representation of a steeply varying function of two variables when the samples are drawn deterministically.

While the estimated frequency of the second mode seems to converge nicely to a steady value as $n_{\max}$ rises at fixed $N$, our results show significantly greater variation with $N$ than those for the dominant mode. Our data indicate that the frequency of this mode might be $0.43 + 0.16i$ with an uncertainty of perhaps 10% in each part. The main reason for the greater uncertainty is that this mode is outgrown by the dominant mode and therefore has a much lower signal-to-noise.

### 3.3. *Grid code results*

As expected, the bias caused by a local smoothing kernel causes the eigenfrequency obtained using the grid code to depend strongly on the softening length used. Figure 4 shows results from simulations with various softening lengths obtained using 120K particles, a grid having 128 azimuthal and 85 radial nodes, and a time step of $0.05 \, (GM/a^3)^{1/2}$, where $M$ and $a$ are defined by (11). Extensive tests showed that variations of grid size and time step changed the estimated frequency by no more than the errors shown, and changes in $N$ had only a slightly larger effect. Despite the suppressed zeros in the vertical scales of these plots, it is clear that both parts of the eigenfrequency are strongly affected by even moderate softening. We tried experiments with a still smaller softening length ($\epsilon = 0.01a$), but the resulting data seemed no longer to be consistent with a few exponentially growing modes. As this problem is made worse by adopting a finer grid, it seems reasonable to attribute the behavior to a more incoherent break-up of the quiet start when the relaxation rate becomes comparable to the growth rate of the mode.

Extrapolation of the main trend in Figure 4 to zero softening indicates an eigenfrequency close to, but somewhat below, Kalnajs's predicted value. The trends are clearly not linear, so a simple linear extrapolation does not yield a fair estimate, although the disagreement in this case is no worse than $\sim 10\%$. We do not try to show that a more sophisticated extrapolation rule could do better, since the result would carry conviction only if, as here, we knew the answer. Thus the grid method yields intrinsically less certain results, unless non-zero softening is required (*cf.* Sellwood & Athanassoula 1986).

### 4. DISCUSSION

The results of the previous section indicate that the SFP method is the best available $N$-body technique for the particular problem of the linear growth of small-amplitude instabilities in collisionless equilibria. This claim is based upon a comparison between our SFP code and a polar grid code on a standard linear stability problem for a thin disk; no other code is likely to be competitive for such a problem. We showed that our SFP code, though somewhat slower than the grid code, produced results in perfect agreement with the linear theory prediction for a continuum disk. Individual results from the grid code are of similar quality, but are biased by the softening kernel; we can therefore estimate the zero softening result only by extrapolation from several experiments with differing softening lengths, making the grid code less precise, and reversing its speed advantage. These strengths of the SFP method are likely to carry over to other stability studies — both of disks and, with an appropriate basis and quiet start, of spheroidal systems.

However, for more general (non-equilibrium) studies that require many more radial functions and Fourier harmonics, the SFP method becomes more expensive and its weaknesses may be, if anything, yet more severe than those of grid codes. The most serious is bias and/or lack of versatility; in order to obtain reasonable performance with a moderate number of functions, the basis set must be well suited to the problem. Different mass distributions may require different bases; a new set can be tailored for a particular problem (Qian 1993; Saha 1993; Earn 1995), but not without effort. A non-adaptive grid code may have difficulty following the evolution when the mass distribution undergoes substantial rearrangement, but the problem seems worse for an SFP code; the basis appropriate at the outset is likely to become inappropriate and may well require additional functions if the later evolution of the model is not to be strongly influenced by the choice of basis.

Our ability to demonstrate relative strengths and weaknesses of different $N$-body methods indicates the importance of choosing a code suited to the problem at hand. Once particle noise on large spatial scales is suppressed by a quiet start, differences between the performance of codes can be seen, in contrast to the conclusions of Hernquist & Barnes (1990) and Hernquist & Ostriker (1992). Unfortunately, quiet starts are appropriate only for initial equilibria and are irrelevant for wide classes of other problems addressable by $N$-body methods.

We wish to thank James Binney and Agris Kalnajs for helpful comments on the original draft of this paper and David Merritt for useful discussions. This work was supported by NSF grant AST-93/18617 and by NASA theory grant NAG 5-2803. DE was supported by a UK Commonwealth Scholarship and a Lady Davis Postdoctoral Fellowship.

## APPENDIX

We give here the formulae for the radial parts of the Abel-Jacobi functions introduced in (10). The original expressions of Kalnajs (1976a) contain a few typographical errors which have been corrected below. Our expressions contain a further factor of $(2\pi)^{-1/2}$ because we require the full Abel-Jacobi functions, not the radial parts alone, to be biorthonormal. If we let

$$\mathcal{P}(k,m,n) = \left[\frac{(2k+m+2n+\tfrac{1}{2})\Gamma(2k+m+n+\tfrac{1}{2})\Gamma(m+n+\tfrac{1}{2})}{\Gamma(2k+n+1)\Gamma^2(m+1)\Gamma(n+1)}\right]^{\tfrac{1}{2}},$$

$$\mathcal{S}(k,m,n) = \frac{\Gamma(k+1)}{\pi\,\Gamma(2k+1)\Gamma(k+\tfrac{1}{2})}\left[\frac{(2k+m+2n+\tfrac{1}{2})\Gamma(2k+n+1)\Gamma(2k+m+n+\tfrac{1}{2})}{\Gamma(m+n+\tfrac{1}{2})\Gamma(n+1)}\right]^{\tfrac{1}{2}},$$

and write $r$ for $r/r_0$, then

$$\phi_{mn}(r) = -\frac{GM}{r_0}\mathcal{P}(k,m,n)\,r^m \sum_{i=0}^{k}\sum_{j=0}^{n}\frac{(-k)_i(m+\tfrac{1}{2})_i(2k+m+n+\tfrac{1}{2})_j(i+m+\tfrac{1}{2})_j(-n)_j}{(m+1)_i(1)_i(m+i+1)_j(m+\tfrac{1}{2})_j(1)_j}\,r^{2i+2j},$$

$$\sigma_{mn}(r) = (-1)^n\frac{M}{r_0^2}\mathcal{S}(k,m,n)\,(1-r^2)^{k-1/2}r^m\sum_{j=0}^{n}\frac{(2k+m+n+\tfrac{1}{2})_j(k+1)_j(-n)_j}{(2k+1)_j(k+\tfrac{1}{2})_j(1)_j}(1-r^2)^j,$$

where $(n)_j$ is the Pochhammer symbol for ratios of factorials (e.g., Arfken 1985, p. 749). We follow Kalnajs's (1976a, §IV) useful advice for the numerical computation of $\phi_{mn}(r)$ and $\sigma_{mn}(r)$ (see also the discussion of efficiency in §2.3).

When coding the basis functions it is usually more convenient to use a real azimuthal basis ($\sin m\theta$ and $\cos m\theta$) rather than the complex basis ($e^{im\theta}$) used in (10). Note that in this case, normalization of the $m \neq 0$ functions requires an extra factor $\sqrt{2}$.

We have struggled somewhat with notation, since the three papers by Kalnajs that are most relevant to us each give different meanings to the same symbols. We have adopted the notation of Kalnajs (1976a). Kalnajs (1972) gave the same meaning to $m$, but $n$ was there the highest power of $r$ in the Legendre polynomial, rather than the number of radial nodes in the function (for $k = 0$ the two indices are related by $n_{1972} = 2 n_{1976} + m$). Kalnajs (1976b and 1978) uses $m$ as an index in the distribution function, which is not related to the azimuthal wave number. Following Sellwood and Athanassoula (1986), we have denoted this by $m_{\rm K}$.

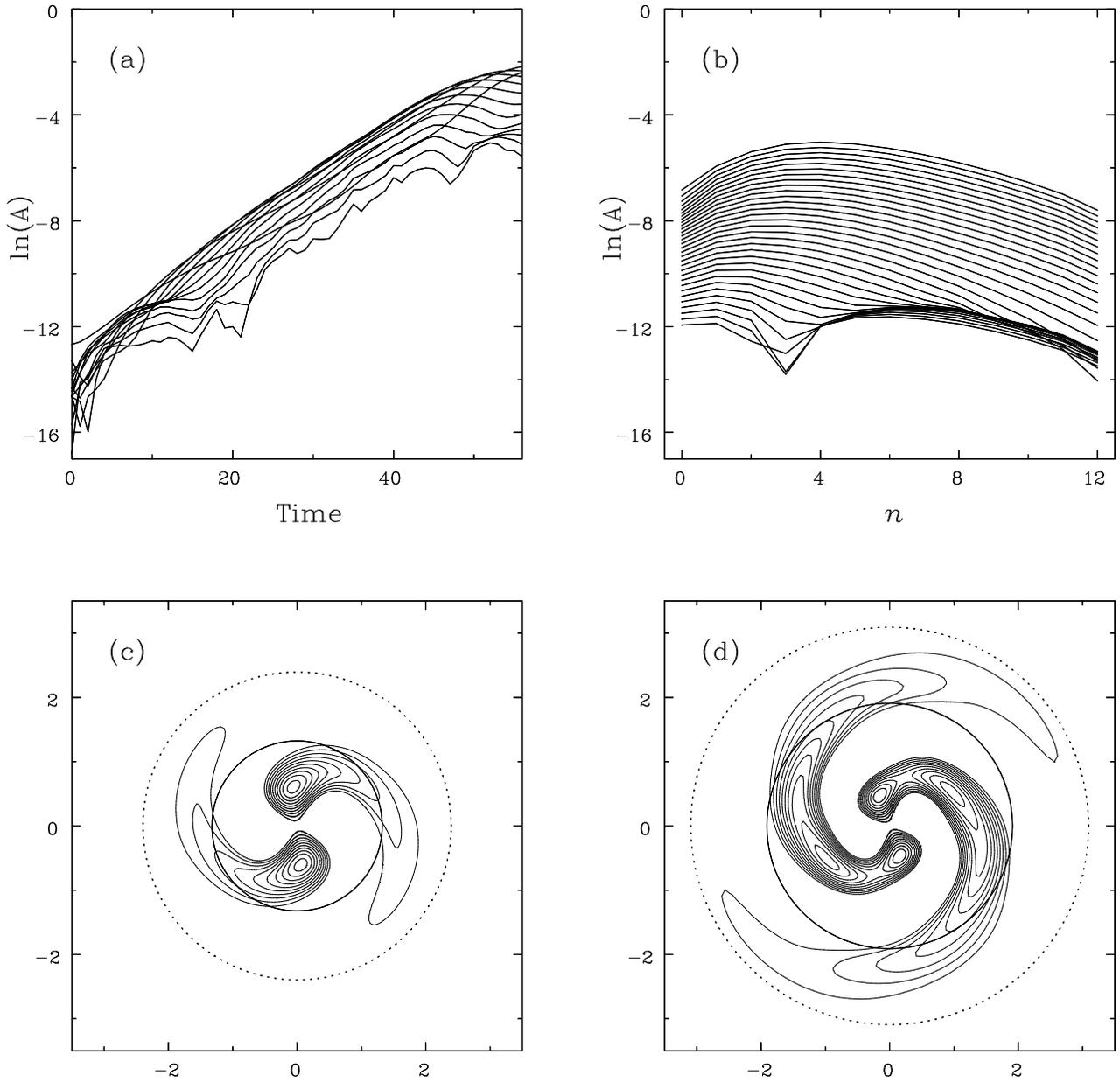

**Figure 1:** The estimation of linear modes. The data here are from an SFP simulation of the isochrone/12 disk with 120K particles, $m = 2$ active only and $n_{\max} = 12$. The time-step was $\Delta t = 0.05$ and the simulation was terminated at $t_{\rm end} = 60$ in natural units ($G = M = a = 1$). (a) The raw data: the amplitudes of each basis member as a function of time. (b) The fitted data displayed to show the two growing modes beating against each other; this fit is made from the subset of the raw data judged to correspond to linear growth (times 5 to 35 in natural units). (c) The shape of the dominant mode (equally spaced density contours of the positive part only). (d) The shape of the secondary mode.

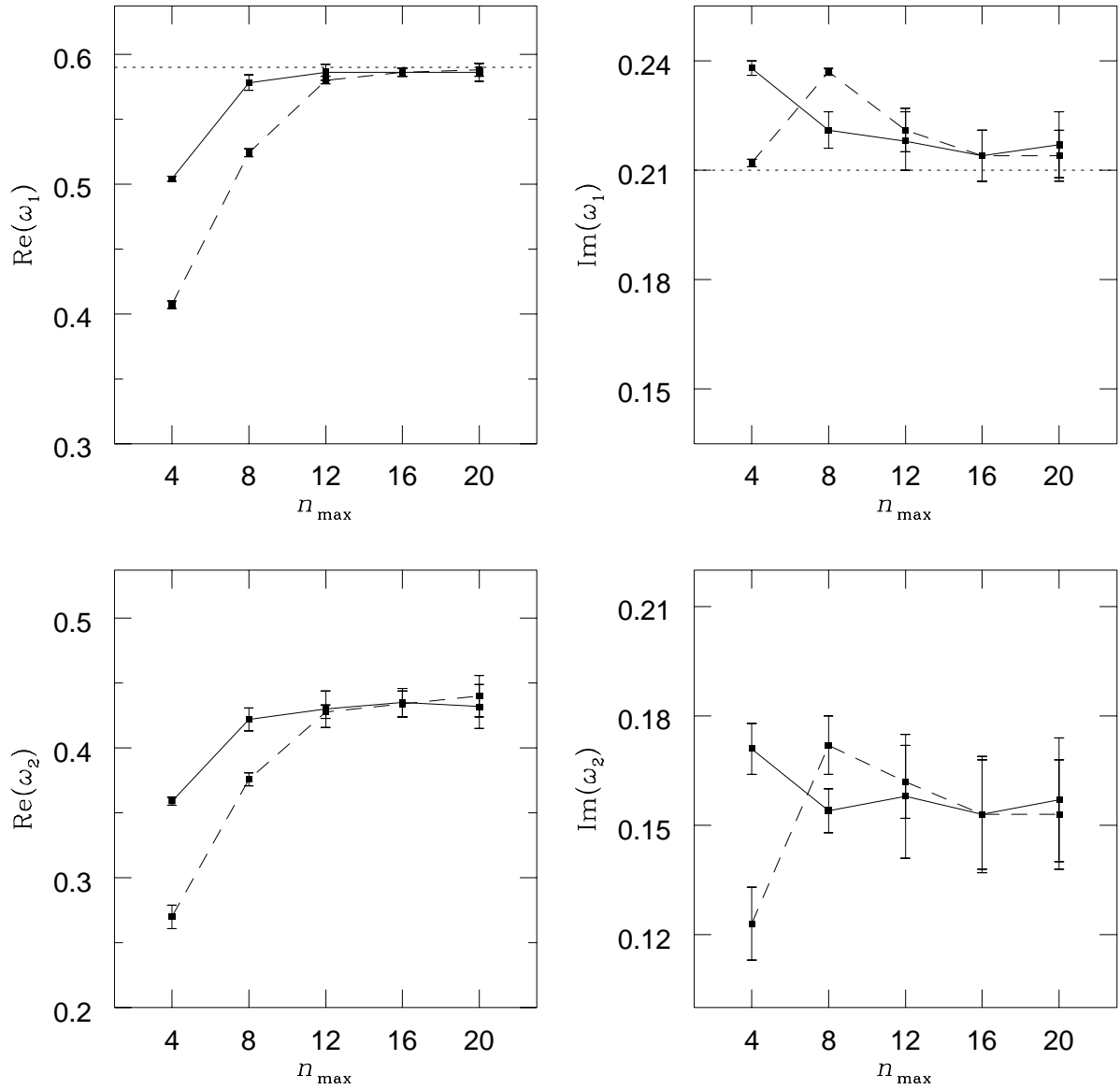

**Figure 2:** Eigenfrequencies obtained by the SFP method for modes of the isochrone/12 disk using 120K particles. The upper panels show the variation with $n_{\max}$ of the pattern speed and growth rate of the dominant $m = 2$ mode. Results using the $k = 7$ Abel-Jacobi basis are joined by solid lines and dashed lines join results from the $k = 2$ set. Horizontal dotted lines show the eigenfrequency, $m\Omega_{\rm p} + \gamma i = 0.59 + 0.21i$, predicted by Kalnajs. The lower panels show the results for the secondary $m = 2$ mode.

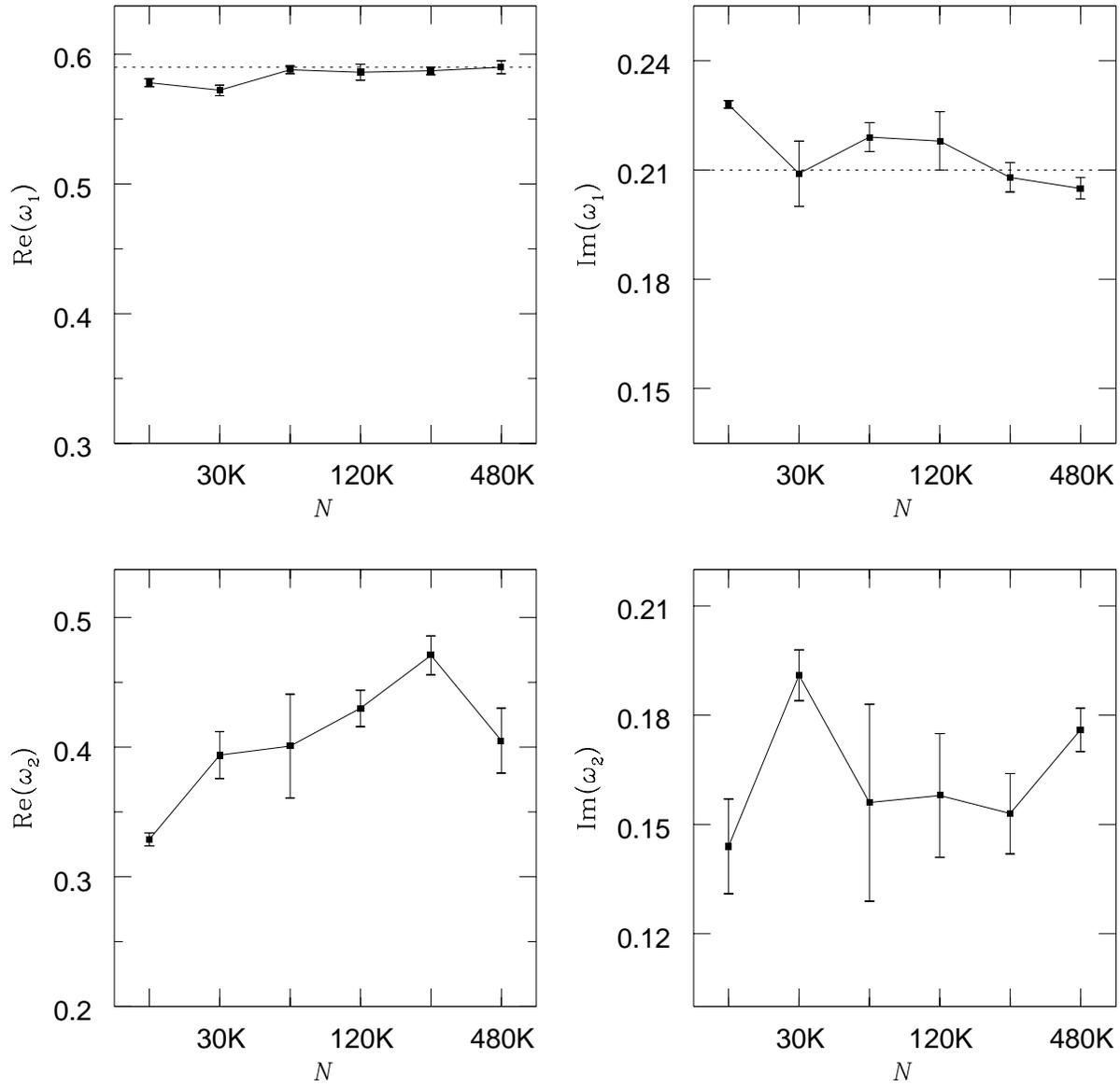

**Figure 3:** The relationship between $m = 2$ eigenfrequencies and the number of particles $N$, from SFP simulations of the isochrone/12 disk with $n_{\max} = 12$. $N$ varies from 15K to 480K (the horizontal scale is logarithmic). As in Figure 2, the upper panels correspond to the dominant mode and the lower panels show results for the secondary mode.

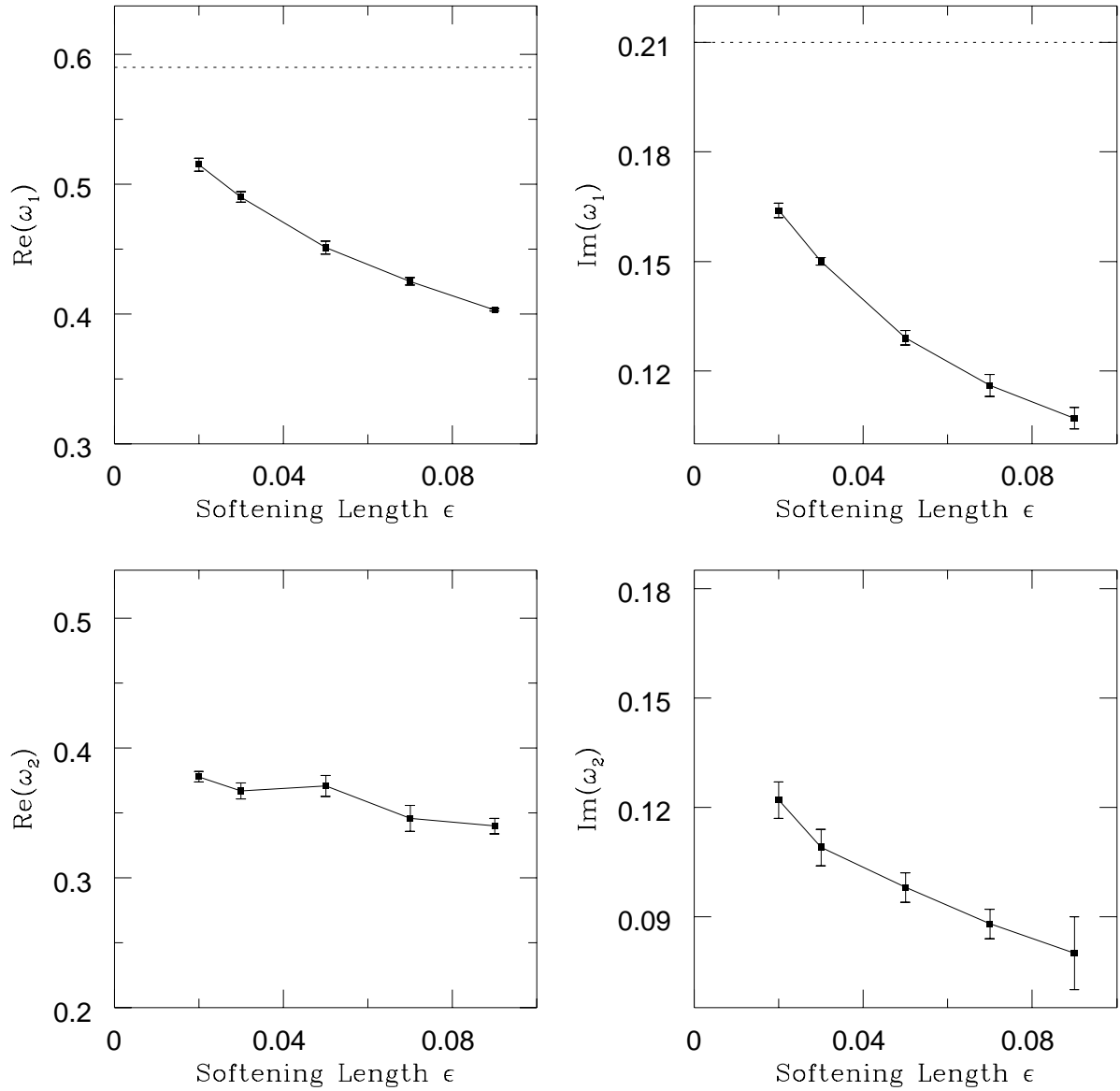

**Figure 4:** Eigenfrequencies obtained with the polar grid code for the $m = 2$ modes of the isochrone/12 disk, with 120K particles and various softening lengths (in units of the disk scale length $a$).